\documentclass[usenatbib]{mn2e}
\usepackage{psfig}

\def\gsim{\mathrel{\raise0.35ex\hbox{$\scriptstyle >$}\kern-0.6em
\lower0.40ex\hbox{{$\scriptstyle \sim$}}}}
\def\lsim{\mathrel{\raise0.35ex\hbox{$\scriptstyle <$}\kern-0.6em
\lower0.40ex\hbox{{$\scriptstyle \sim$}}}}
\def\gs{\mathrel{\raise0.35ex\hbox{$\scriptstyle >$}\kern-0.6em
\lower0.40ex\hbox{{$\scriptstyle \sim$}}}}
\def\ls{\mathrel{\raise0.35ex\hbox{$\scriptstyle <$}\kern-0.6em
\lower0.40ex\hbox{{$\scriptstyle \sim$}}}}

\def\kms{\,\hbox{km}\,\hbox{s}^{-1}}
\def\Msol{\mathrel{\rm M_{\odot}}}
\def\Msolyr{\mathrel{\rm M_{\odot}\,\hbox{yr}^{-1}}}
\def\Wm2{\,\hbox{W}\,\hbox{m}^{-2}}

\begin{document}

\title[Optical \& Near-IR IFU Spectroscopy of a z=5 Lensed
Proto-Galaxy] {Resolved Spectroscopy of a Gravitationally Lensed
  L$^{*}$ Lyman-break Galaxy at z$\sim$5}

\author[Swinbank et al.]{
\parbox[h]{\textwidth}{
A.\,M.\ Swinbank$^{1\, *}$,
R.\,G.\ Bower$^1$,
Graham\ P.\ Smith$^{2,4}$,
R.\,J.\ Wilman$^{1}$,
\\
Ian Smail$^1$,
R.\,S.\ Ellis$^2$,
S.\,L.\ Morris$^1$,
\& J.-P.\ Kneib$^{3}$}
\vspace*{6pt} \\
$^1$Institute for Computational Cosmology, Department of Physics, Durham University, South
Road, Durham DH1 3LE, UK \\
$^2$California Institute of Technology, MC 105-24, Pasadena, CA 91125, USA \\
$^3$Laboratoire d'Astrophysique de Marseille, Traverse du Siphon - B.P.8 13376, Marseille Cedec 12, France\\
$^4$School of Physics and Astronomy, University of Birmingham, Edgbaston, Birmingham, B15 2TT \\
$^*$Email: a.m.swinbank@dur.ac.uk \\
}
\setcounter{footnote}{0}

\maketitle

\begin{abstract}
  We exploit the gravitational potential of a massive, rich cluster at
  $z=0.9$ to study the internal properties of a gravitationally lensed
  galaxy at $z$=4.88.  Using high resolution {\it HST} imaging together
  with optical (VIMOS) and near-infrared (SINFONI) Integral Field
  Spectroscopy we have studied the rest-frame UV and optical properties
  of the lensed galaxy seen through the cluster RCS0224-002.  Using a
  detailed gravitational lens model of the cluster we reconstruct the
  source-frame morphology on 200pc scales and find an $\sim$L$^{*}$
  Lyman-break galaxy with an intrinsic size of only
  2.0$\times$0.8\,kpc, a velocity gradient of $\lsim$60$\kms$ and an
  implied dynamical mass of 1.0$\times$10$^{10}$$\Msol$ within 2\,kpc.
  We infer an integrated star-formation rate of just 12$\pm$2$\Msolyr$
  from the intrinsic [O{\sc ii}]$\lambda$3727 emission line flux.  The
  Ly$\alpha$ emission appears redshifted by +200$\pm$40$\kms$ with
  respect to the [O{\sc ii}] emission.  The Ly$\alpha$ is also
  significantly more extended than the nebular emission, extending over
  11.9$\times$2.4\,kpc. Over this area, the Ly$\alpha$ centroid varies
  by less than 10$\kms$.  We model the asymmetric Ly$\alpha$ emission
  with an underlying Gaussian profile with an absorber in the blue wing
  and find that the underlying Ly$\alpha$ emission line centroid is in
  excellent agreement with the [O{\sc ii}] emission line redshift.  By
  examining the spatially resolved structure of the [O{\sc ii}] and
  Ly$\alpha$ emission lines we investigate the nature of this system.
  The model for local starburst galaxies suggested by Mass-Hesse et
  al.\ (2003) provides a good description of our data, and suggests
  that the galaxy is surrounded by a galactic-scale bi-polar outflow
  which has recently burst out of the system.  The outflow, which
  appears to be currently located $\gsim$30\,kpc from the galaxy, is
  escaping at a speed of upto $\sim$500\,$\kms$.  Although the mass of
  the outflow is uncertain, the geometry and velocity of the outflow
  suggests that the ejected material is travelling far faster than
  escape velocity and will travel more than 1\,Mpc (comoving) before
  eventually stalling.
\end{abstract}


\begin{figure*}
  \centerline{
    \psfig{file=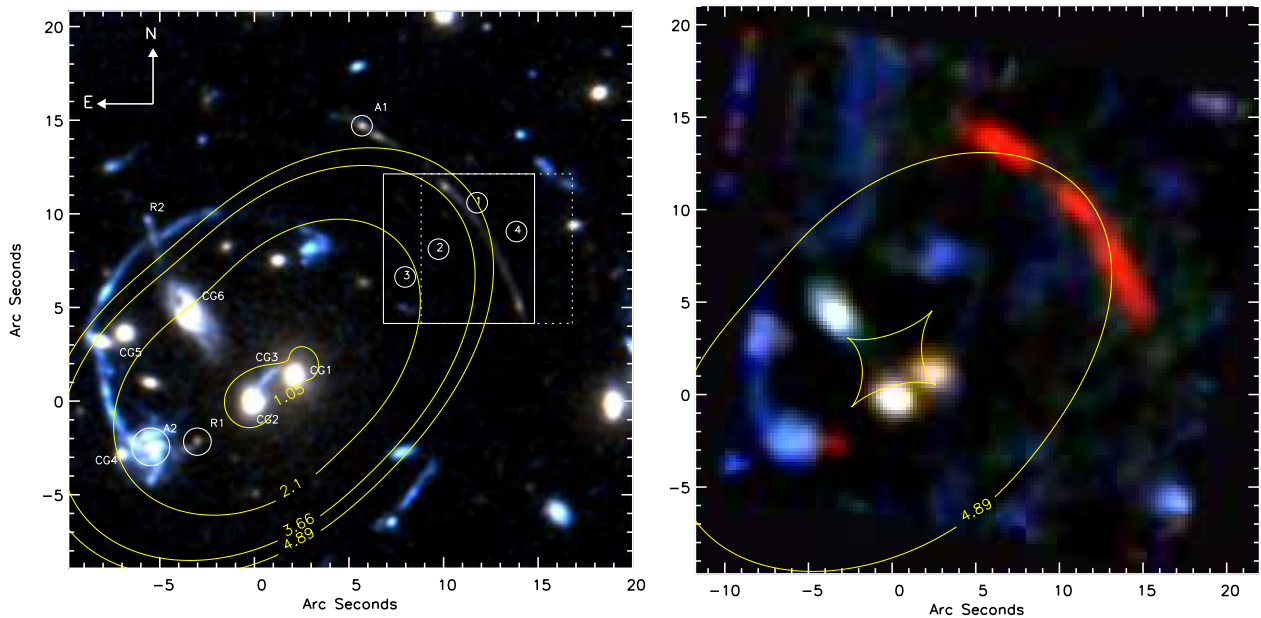,width=6.6in,angle=90}}
  \caption{{\it Left:} True colour {\it HST} $VI$-band image of the
    core of the lensing cluster RCS0224-002 at $z$=0.78.  The contours
    mark the high-redshift critical curves (curves of infinite
    magnification) from the gravitational lens model described in \S3.
    We also overlay the field of view of the SINFONI IFU (shown by the
    white box) which was used to map the [O{\sc ii}]$\lambda$3727
    emission.  The cluster galaxies which we are able to
    spectroscopically identify are labelled CG1-6.  R1 is the radial
    counter-image of the $z$=4.88 arc and R2 is the $z$=1.05 radial arc
    from Sand et al.  (2005).  Serendipitous background galaxies are
    labelled A1 and A2 (VIMOS) and 1--4 (SINFONI) (see
    Appendix~\ref{sec:serendip}).  {\it Right:} $VR$($I$+Ly$\alpha$)
    colour image of the cluster core generated from the VIMOS IFU
    datacube.  The inner and outer curves show the $z$=4.88 caustic and
    critical curves respectively.  The center of the cluster (0,0) is
    at $\alpha=$02:24:34.255 $\delta=$-00:02:32.39 (J2000) and we have
    rotated and aligned the {\it HST} and VIMOS data such that North is
    up and East is left in both panels.}
\label{fig:hst+vimos_col}
\end{figure*}

\section{Introduction}
One of the most important observational breakthroughs in recent years
was the discovery that a significant fraction of high-redshift galaxies
are surrounded by ``superwinds''
\citep[e.g.][]{Pettini02,Shapley03,Bower04,Wilman05} -- starburst
and/or AGN driven outflows which expel gas from the galaxy potential,
hence playing no further role in the star-formation history of the
galaxy.  This phenomenon is beginning to be understood by theorists as
the missing link in galaxy formation models which are otherwise unable
to match the shape and normalisation of the luminosity function
\citep{Benson03,Baugh04}.  These feedback processes may also offer
natural explanation as to why only 10\% of baryons cool to form stars
(the Cosmic Cooling Crisis; \citealt{White78,Balogh01_ccc}).

However, important questions remain unanswered.  Evidence for these
superwinds is usually based on observations which compare the nebular
emission line properties with the rest-frame UV emission and absorption
lines (such as Ly$\alpha$, H$\alpha$ and UV ISM absorption lines;
e.g.\, \citealt{Erb03}).  Velocity offsets of several hundred km/s have
been measured, suggestive of large scale outflows comparable to
starburst driven winds often observed in low-redshift Ultra-Luminous
Infra-Red Galaxies (ULIRGs) in the local Universe \citep{Martin05}.
However, the current data lack spatial information, which is vital if
we are to understand if material escapes into the Inter-Galactic Medium
(IGM) or whether the outflowing material eventually stalls, fragments
and drains back onto the galaxy, potentially disrupting the disk and
causing further bursts of star-formation.

The key to resolving these issues is to identify similar features in
the spatially resolved spectra of distant, young galaxies.  However, at
the redshifts where these feedback processes are at their peak activity
(above $z\sim$2, i.e. when galaxies were most rapidly forming stars),
the immense luminosity distances mean that the emission line fluxes of
these distant galaxies are dramatically reduced.  This problem is
compounded by the fact that the nebular emission lines (such as [O{\sc
  ii}], [O{\sc iii}] and H$\alpha$) are redshifted into the
near-infrared, where the sky background is an order of magnitude
brighter than in the optical.  Coupled with the fact that galaxies at
these redshifts typically have disk scale lengths of only a few kpc,
even with adaptive optics assisted observations on eight to ten meter
telescopes, only a handful of independent pixels can be recovered from
a distant galaxy.

By necessity therefore, above $z\sim2$ most detailed studies of
individual galaxies have concentrated on the most luminous (and
therefore usually very active and massive) systems
\citep{Bower04,Wilman05,Swinbank05b}, with very few ``normal'' galaxies
being studied in detail beyond $z=2$
\citep{Pettini02,Pettini00,Teplitz00,ForsterSchreiber06,Genzel06}.  One
way to overcome this problem is to use the deep potential of massive
galaxy clusters to boost the flux and sizes of images of distant
galaxies which serendipitously lie behind them.  This natural
magnification provides the opportunity to study intrinsically faint
high redshift galaxies with a spatial resolution that cannot be
attained via conventional observations
\citep{Smail96,Franx97,Teplitz00,Ellis01,Campusano01,Smith02,Swinbank03,Kneib04b,Swinbank06a}.

The natural magnification caused by the gravitational lens allows us to
spatially resolve the morphologies and internal dynamics of galaxies in
a level of detail far greater than otherwise possible.  For the galaxy
discussed in this paper, the amplification factor of sixteen
corresponds to an increase of over three magnitudes and makes it
possible to study the galaxy on independent spatial scales of
$\sim$200pc at $z=$5.  In contrast, without a gravitational lens, 1$''$
corresponds to 6.5\,kpc at $z$=4.88.

In this paper we present a VLT/IFU study of the $z=4.88$
gravitationally lensed galaxy behind the core of the rich lensing
cluster RCS0224-002.  This galaxy cluster was identified as having a
significant concentration of red galaxies in the inner regions by
\citet{Gladders02}.  Follow-up spectroscopy confirmed a redshift for
the cluster of 0.773$\pm$0.0021.  The ground based imaging from
\citet{Gladders02} shows several tangential arc-like structures, many
of which resemble images of lensed, background galaxies.  One of the
most striking of these is a multiply imaged arc approximately 15$''$ to
the North-East of the Brightest Cluster Galaxy (BCG) (see
Fig.~\ref{fig:hst+vimos_col}). Follow-up spectroscopy by
\citet{Gladders02} yielded a redshift of $z$=4.88 from identification
of a strong Ly$\alpha$ emission line at 7148\AA.

We concentrate on the dynamics and star-forming properties of the
$z$=4.88 arc observed through the rest-frame UV spectra with the VIMOS
IFU and through the nebular ([O{\sc
  ii}]$\lambda\lambda$3726.1,3728.8\AA) emission line doublet observed
with the SINFONI IFU in the near-infrared. The IFU data provide a
two-dimensional map of the galaxy's properties in sky co-ordinates.  In
order to interpret this map we must correct for the magnification and
distortion caused by the lensing potential using a mass model for the
cluster lens.  This model is constrained by requiring it to account for
the positions and redshifts of the gravitationally-lensed arcs in the
cluster.  The lensing model then allows us to determine the source
plane morphology and geometry of the continuum and line emission we
use.

The structure of this paper is as follows.  In \S2, we present the data
on which this paper is based: a combination of {\it Hubble Space
  Telescope (HST)}, ground-based imaging, and integral field
spectroscopy obtained with the SINFONI and VIMOS IFUs on the VLT.  In
\S3 we present the source-plane properties of the arc and the spatially
resolved spectroscopy and in \S4 and \S5 we present our discussion and
conclusions respectively.  Through-out this paper we use the Vega
magnitude system and assume a cosmology with $H_{0}=72\kms$,
$\Omega_{0}=0.3$ and $\Lambda_{0}=0.7$.


\begin{figure}
  \centerline{
    \psfig{file=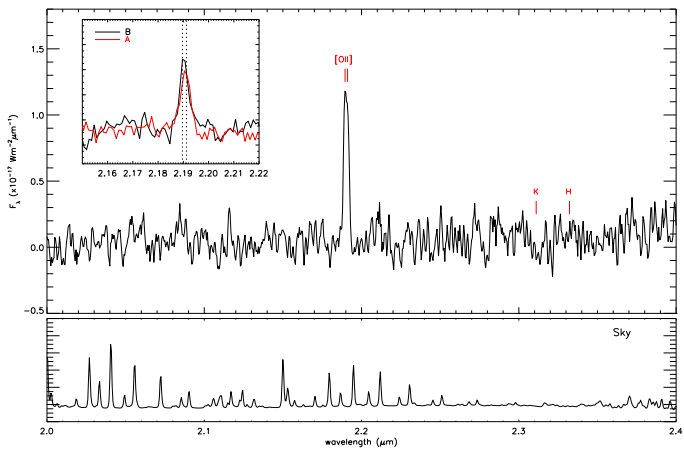,width=3.7in,angle=90}}
  \caption{Collapsed, one-dimensional spectrum of the arc from our
    SINFONI IFU observations.  The [O{\sc ii}] emission line has a
    redshift of 4.8757$\pm$0.0005 and a line width of
    $\sigma$=180$\pm$30$\kms$.  The inset shows the renormalised [O{\sc
      ii}] emission lines for the two regions (A\&B) shown in Figure~4.
    We also show a sky-spectrum (below) for comparison (scaled in flux
    for clarity).}
\label{fig:sinf_1dspec}
\end{figure}

\begin{figure}
  \centerline{
    \psfig{file=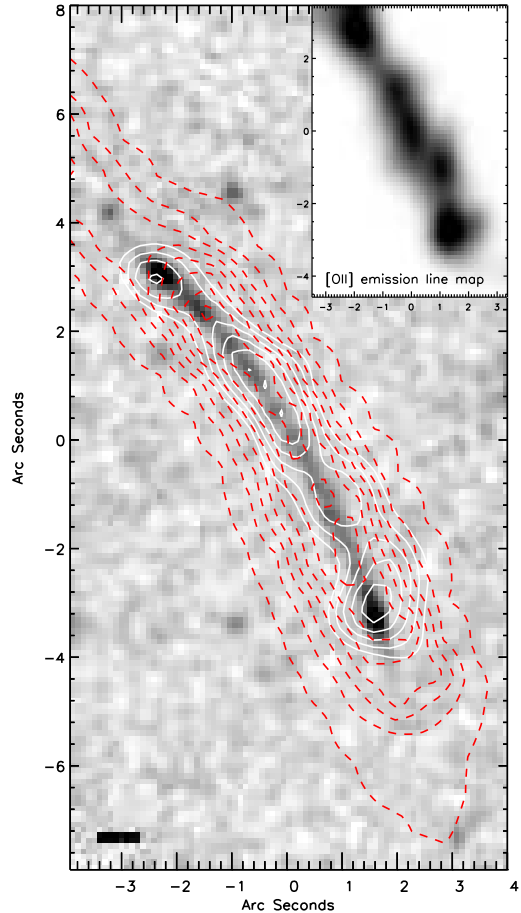,width=2.7in,angle=0}}
  \caption{Image-plane observations of the $z$=4.88 arc: {\it HST}
    continuum (greyscale), Ly$\alpha$ emission (red contours) and
    [O{\sc ii}] emission (white contours).  The image shows that the
    Ly$\alpha$ emission is much more extended than either the [O{\sc
      ii}] or continuum morphology.  The solid bar in the top right
    hand corner represents the 0.8$''$ seeing for both the VIMOS and
    SINFONI IFU observations.{\it Inset:} Continuum subtracted,
    narrow-band image around the redshifted [O{\sc ii}]$\lambda$3727
    emission of the arc from our SINFONI IFU observations.  The highest
    surface brightness components in both the optical image and the
    [O{\sc ii}] emission line map are well matched, though the [O{\sc
      ii}] emission line map also shows that the highly sheared
    component has more structure than evident in the optical imaging.
    {\it Right:} }
\label{fig:sky_overlays}
\end{figure}

\begin{figure*}
  \centerline{
    \psfig{file=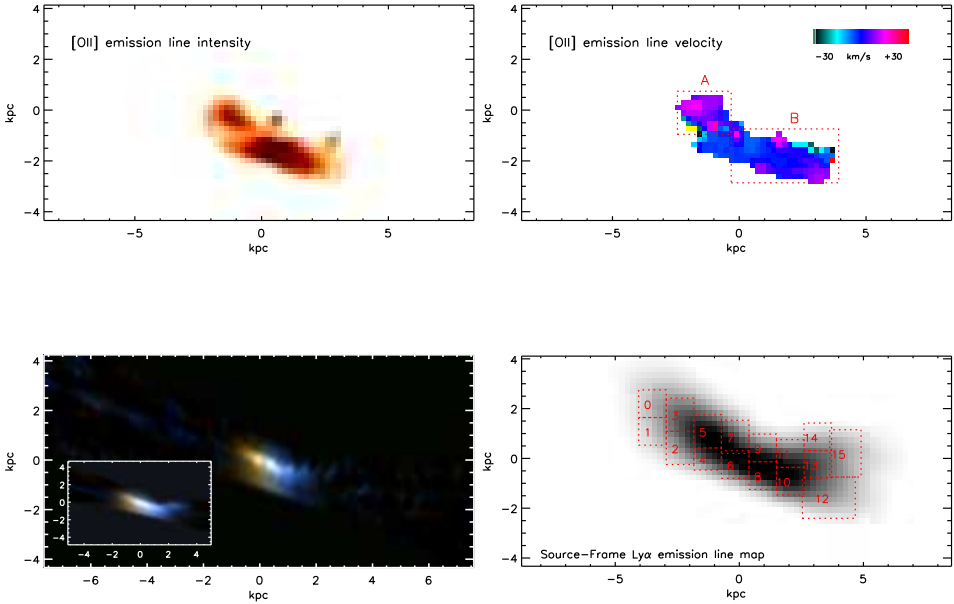,width=5in,angle=90}}
  \caption{Source-plane observations of the $z$=4.88 galaxy from the {\it HST}, 
    VIMOS and SINFONI observations. {\it Top Left:} [O{\sc ii}]
    emission line intensity of the galaxy (dark regions represent
    regions of highest intensity). {\it Top Right:} [O{\sc ii}]
    emission line velocity structure of the galaxy which shows a
    maximum velocity shift of 60$\pm$20$\kms$ along the long axis of
    the galaxy.  {\it Bottom Left:} Reconstructed true colour {\it HST}
    $VI$ image of the $z$=4.88 arc.  {\it Inset:} Reconstructed {\it
      HST} image after a smoothing scale of 0.8$''$ has been applied to
    the sky-plane image.  {\it Bottom Right:} Reconstructed Ly$\alpha$
    emission line map of the $z$=4.88 arc.  The boxes show regions from
    which the spectra in Figure~6 were extracted.}
\label{fig:sourceall}
\end{figure*}


\section{Observations and Data Reduction}
\label{sec:obs+redux}

\subsection {{\it HST} imaging}
\label{sec:HSTobs}

{\it HST} WFPC2 $I_{814}-$ and V$_{555}-$band observations of the
lensing cluster RCS0224-002 were obtained from the {\it HST} public
archive\footnotemark.  The $I_{814}$ and $V_{555}$-band observations
were 10.5 and 8.4\,ks respectively and the data were reduced using the
standard {\sc stsdas} package in {\sc iraf}.  The resulting image
(Fig.~\ref{fig:hst+vimos_col}) covers the brightest central cluster
galaxies and the target arc at a resolution of 0.0996$''$/pixel.  From
the {\it HST} imaging the $z$=4.88 arc is red
($V_{606}-I_{814}$=1.7$\pm$0.1) with an average surface brightness of
$\mu_{I_{814}}$=25.0 and an integrated magnitude of $I$=22.2.

\footnotetext{PID: 9135; Obtained from the Multimission Archive at the
  Space Telescope Science Institute (MAST).  STScI is operated by the
  Association of Universities for Research in Astronomy, Inc., under
  NASA contract NAS5-26555. Support for MAST for non-HST data is
  provided by the NASA Office of Space Science via grant NAG5-7584 and
  by other grants and contracts.}

\subsection{VIMOS Integral Field Spectroscopy}
\label{sec:vimos_ifs}

The $z$=4.88 arc was observed with the VIMOS IFU \citep{LeFevre03} for
a total of 43.2ks (split into 16$\times$2700 second exposures) between
2004 December 16 and 2005 December 12 in $\lsim$0.6$''$ seeing and
photometric conditions.  We used the MR-orange grism which results in a
field of view of 27$''$$\times$27$''$ at 0.67$''$ per pixel and a
spectra resolution of $\lambda$/$\Delta\lambda\sim$1100 at 7000\AA.  To
reduce the data we used the VIMOS {\sc esorex} pipeline which extracts,
wavelength calibrates and flat-fields the data.  The final datacubes
cover a wavelength range of 5000-11000\AA\ and has a resolution of
6.6\AA\ FWHM (measured from the width of the sky-lines at a wavelength
of 7150\AA).  In all future sections, line widths are deconvolved for
instrumental resolution.  Flux calibration was achieved by using
observations of ESO standard stars.  These observations were taken
either immediately before, or immediately after the science
observations and were reduced in an identical manner.  Since there are
no point sources in our datacubes, we measure the seeing from the
standard stars in the collapsed datacubes and derive typical seeing
measurement of $\sim$0.8$''$ at a typical airmass 1.1.  During the
standard star observation, the ESO seeing monitor registered median
seeing of 0.7$''$ -- hence we conservatively assume that the VIMOS IFU
observations were taken in 0.8$''$ seeing.

After reducing each of the observations to a datacube, we construct and
apply a master flat-field in order to correct for the fringing effects
which are particular prominent above 8000\AA.  In each of the
(spatially dithered) datacubes we construct a broad-band image by
collapsing the datacube in wavelength between 6000\AA\ and 7500\AA\ and mask out the the brightest objects.  We then apply this mask at each
wavelength and stack the resulting datacube (without spatial offsets
and ignoring masked pixels).  This produces a ``master-flatfield''
which is then divided into each of the science observations to reduce
the effects of the fringing.  These individual datacubes are then
sky-subtracted by masking the brightest objects in the individual
frames and constructing a sky-spectrum by collapsing the datacube in
both spatial dimensions (since the four quadrants of the VIMOS IFU have
slightly different throughput's, we note that the sky-subtraction was
performed quadrant-by-quadrant).  To build the final datacube, we
create broad band images from each datacube and cross correlate these
in order to spatially align and create the final mosaic.  The final
datacube is created using an average with a 3$\sigma$ clip to reject
cosmic rays.

In Fig.~\ref{fig:hst+vimos_col} we show a colour image of the cluster
and arc, generated from the datacube.  We generate three colours from
the datacube by collapsing the datacube between 5000-6000\AA,
6000-7000\AA\ and 7000-8000\AA\ as blue, green and red respectively.
Alternatively, the data-cube can be viewed as a sequence of wavelength
slices to make an animation (this can be viewed at:\\
http://star-www.dur.ac.uk/$\sim$ams/lensing/RCS0224$\_$z5/ )

\begin{figure}
  \centerline{
    \psfig{file=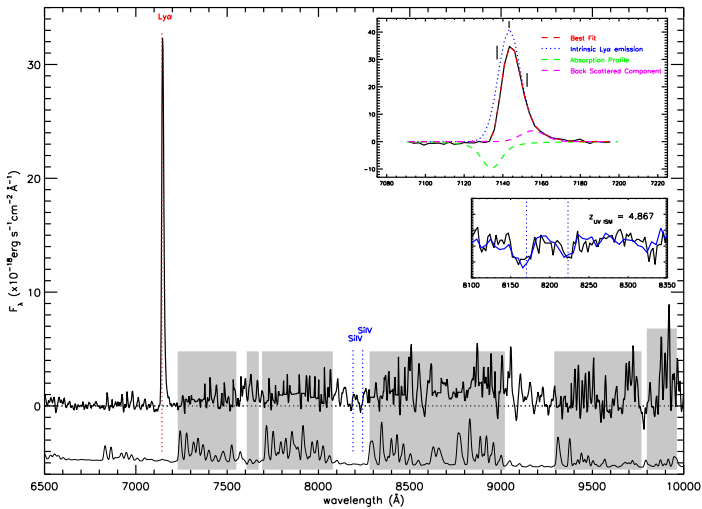,width=3.8in,angle=90}}
  \caption{Collapsed, one-dimensional spectrum of the arc from our
    VIMOS IFU observations.  The dashed vertical lines show the
    expected position of the UV emission and absorption lines for the
    systemic redshift of $z=$4.8757 (as measured from the [O{\sc ii}]
    emission line).  The Ly$\alpha$ emission appears redshifted with
    respect to the systemic velocity, and has an asymmetric line
    profile.  We also detect the weak UV-ISM lines of Si{\sc iv}. These
    lie in a region devoid of strong sky emission.  The hashed regions
    show regions of strong sky emission and the horizontal dashed line
    marks a continuum level of zero. {\it Top Inset:} The Ly$\alpha$
    emission line with the model described in \S3.2.2.  The black
    (solid) line shows the observed spectra.  The asymmetric line is
    best fit with an underlying Gaussian emission line profile (blue
    dotted line) combined with a Voigt profile absorber (green dashed
    line) and a third Gaussian which accounts for the red wing of
    emission.  Whilst the observed Ly$\alpha$ emission appears
    redshifted from the systemic velocity, the best-fit underlying
    emission profile has a centroid in excellent agreement with the
    [O{\sc ii}] emission.  {\it Lower Inset:} Spectra around the
    rest-frame UV absorption lines Si{\sc
      iv}$\lambda\lambda$1393.76,1402.8.  We overlay the best-fit
    template (blue) from Shapley et al. (2003) and use this to infer a
    velocity shift of -400$\pm$100$\kms$ from the systemic redshift.  }
\label{fig:vimos_1dspec}
\end{figure}

\subsection{SINFONI IFU Observations}
\label{sinfoni_ifs}

To map the nebular emission line properties of the $z$=4.88 arc, we
used the SINFONI IFU \citep{Eisenhauer03}.  At $z$=4.88 the [O{\sc
  ii}]$\lambda\lambda$3726.1,3728.8 emission line doublet is redshifted
to 2.19$\mu$m and into a stable part of the $K$-band relatively free
from strong OH airglow emission.  The SINFONI IFU uses an image slicer
and mirrors to reformat a field of 10$\times$10$''$ at a spatial
resolution of 0.25$''$/pixel.  We used the HK grating which results in
a spectral resolution of $\lambda$/$\Delta\lambda$=1700 at 2.20$\mu$m
(the sky-lines have a resolution 13\AA\ FWHM at this wavelength) and
covers a wavelength range of 1.451 to 2.463$\mu$m.  In all future
sections, emission line widths are deconvolved for the instrumental
resolution.  To observe the target we used ABBA chop sequences, however
since the arc only filled one half of the datacube, we chopped 2$''$
West to sky (see Fig.~\ref{fig:hst+vimos_col}), thus keeping the object
inside the IFU.  We observed the target for a total of 43.2\,ks (split
into 16$\times$2700 second exposures) between 2005 August 10 and 2005
September 28 in $\lsim$0.6$''$ seeing and photometric conditions.
Individual exposures were reduced using the SINFONI {\sc esorex} data
reduction pipeline which extracts, flatfields, wavelength calibrates
and forms the datacube.  The final datacube was generated by aligning
the individual data-cubes and then combining the using an average with
a 3$\sigma$ clip to reject cosmic rays.  For flux calibration, standard
stars were observed each night during either immediately before or
after the science exposures.  These were reduced in an identical manner
to the science observations.  In Figure~\ref{fig:sinf_1dspec} we show
the one-dimensional spectrum of the arc (generated by collapsing the
datacube over the object in both spatial domains) and in
Figure~\ref{fig:sky_overlays} we also show a narrow-band image of the
arc generated by collapsing the datacube in the wavelength direction
over the [O{\sc ii}] emission line.  Figure~\ref{fig:sky_overlays} also
shows the {\it HST} image of the arc with the contours from the [O{\sc
  ii}] and Ly$\alpha$ overlaid.

\section{Analysis \& Results}
\label{sec:resoved_spectra}

\subsection{Mass Modelling} 
\label{sec:mass_modelling}

Reconstruction of the intrinsic properties of the galaxy at $z{=}4.88$
requires removal of the gravitational magnification from the
observables.  We therefore use the four spectroscopically confirmed
images of the $z{=}4.88$ galaxy to constrain a model of the mass
distribution in the cluster.  These four images comprise the three
images in the tangential arc identified by \citet{Gladders02} and the
fourth image identified via our integral field spectroscopy
(Appendix~\ref{sec:serendip}).  The candidate radial arc at $z{=}1.05$
(A3 in Fig.~\ref{fig:hst+vimos_col}; \citealt{Sand05}) and the object
identified at $z{=}3.66$ adjacent to the $z{=}4.88$ arc (A1 in
Fig.~\ref{fig:hst+vimos_col}) appear not to be multiply-imaged, and so
are not included as constraints.  The configuration of the four
observed images and the lack of strong radial amplification of the
fourth image implies that the fifth image lies very close to the centre
of the cluster mass distribution.  We therefore expect the fifth image
to be strongly de-magnified and thus likely undetectable in the current
data.  We identify the dense knot of emission in each of the four
observed images as being images of the same underlying region of the
galaxy, and use the position of these knots to constrain the model.
Following Appendix~A of \citet{Smith05}, the number of constraints is
therefore $n_{\rm C}{=}6$.

The central region of RCS0224 contains two bright elliptical galaxies:
CG1 and CG2 (Fig.~\ref{fig:hst+vimos_col}) -- we therefore construct a
three component model comprising the cluster-scale mass distribution
(dark matter and gas), CG1 and CG2.  We parametrise all three mass
components as truncated pseudo-isothermal elliptical mass
distributions, each described by the following parameters:
$\{x,y,\epsilon,\theta,r_{\rm core},r_{\rm cut},v_{\rm disp}\}$
\citep{Kassiola93,Kneib96} -- i.e.\ 21 parameters in total, comprising
the center, ellipticity, position angle, core radius and cut-off radius
of each mass component.  To obtain a well-constrained model, (and given
the number of constraints above: $n_{\rm C}{=}6$) we prefer no more
than 6 of these 21 parameters to be free parameters in the lens model.
The geometrical parameters ($x,y,{\epsilon},{\theta}$) describing CG1
and CG2 are therefore matched to their 2-dimensional light
distributions, and their respective $r_{\rm core}$ and $r_{\rm cut}$
parameters are matched to the shape of the light profile of each
galaxy.  Finally, the mass-to-light ratio of each galaxy is fixed by
matching their velocity dispersions to the measured velocity dispersion
(Table~\ref{table:cluster_members}).  Turning to the parameters
describing the cluster-scale mass, we fix both $r_{\rm core}$ and
$r_{\rm cut}$ at $75\,{\rm kpc}$ and $1\,{\rm Mpc}$ respectively.  The
former value is representative of massive clusters, and our decision to
fix this parameter is caused by the non-detection of the central fifth
image of the $z{=}4.88$ arc.  The lack of the fifth image reduces our
ability to constrain the radial shape of the mass distribution within
the tangential critical curve.  We also run models with $r_{\rm
  core}{=}50\,{\rm kpc}$ and $100\,{\rm kpc}$ to confirm that the final
results are not altered by the adopted value of $r_{\rm core}$.  For
completeness, we include the result of this check in the error on the
lens magnification calculated below.  Fixing $r_{\rm cut}{=}1\,{\rm
  Mpc}$ is motivated by the small field of view of the \emph{HST} data
and within those data the small angular scale subtended by the strong
lensing constraints.  We are therefore unable to constrain this
parameter, but our results are insensitive to its value.  In summary,
we therefore fit a model with five free parameters to the data.  The
free and fixed parameters are listed in Table~1.

All of the lens modelling is performed using the {\sc lenstool}
software \citep{Kneib93PhD,Kneib96}, incorporating a Markov Chain
Monte Carlo (MCMC) sampler (Jullo et al.\ 2006, in preparation).  The
model parameters with the lowest ${\rm \chi}^2$ found with this MCMC
approach and $68\%$ confidence intervals around these parameters
values after marginalising over the other four parameters in each
case, are listed in Table~1, and we show the critical curve of the
best fit model in Fig.~\ref{fig:hst+vimos_col}.  The lens model
predicts the fifth image to lie at the center of CG1 and to be very
faint -- $R{\gsim}36$, i.e.\ de-magnified by ${\sim}14$ magnitudes
relative to the tangential arc.  Although the model makes a clear
prediction for the location of this image, it is clearly too faint to
detect with the available data.  We ray-trace the $z{=}4.88$ galaxy
through the family of models within the $68\%$ confidence surface in
the 5-dimensional parameter space to compute the mean, luminosity
weighted magnification of $\mu$=16$\pm$2 (which corresponds to
$\Delta$m=3.0$\pm$0.2 magnitudes).

Accounting for the lensing magnification, the intrinsic magnitude of
this galaxy is $I$=25.2.  In comparison, deep imaging surveys of
Lyman-break galaxies at z$\sim$5 have derived an $L^{*}$ of $i\sim$25.3
\citep{Ouchi04}.  This suggests that the $z=$4.88 galaxy is typical of
the UV continuum selected LBG population at this redshift.  In
Fig.~\ref{fig:sourceall} we show the reconstructed {\it HST} image of
the galaxy.  In the source-plane the galaxy has a FWHM of only
2.0$\times$0.8\,kpc, and indicates a flattened (or bar-like) geometry.
Such an elongated morphology is not unusual for galaxies at these
redshifts: in a recent high-resolution imaging survey of LBGs at
z$>$2.5 with {\it HST}, Navindranath et al. (2006) concluded that upto
50\% of LBGs show evidence for bar-like morphologies (with mean scale
lengths of 1.7--2.0\,kpc).  Thus the galaxy appears typical in terms of
its source plane morphology, brightness and size of the population at
$z\sim5$.

\begin{table*}
{\scriptsize
\begin{center}
{\centerline{\sc Table 1.}}
{\centerline{\sc Gravitational Lens Model Parameters}}
\smallskip
\begin{tabular}{lccccccc}
\hline
\hline
\noalign{\smallskip}
          & ${\Delta}{\rm RA}$ & ${\Delta}{\rm Dec}$ & $\epsilon$         & $\theta$           & $r_{\rm core}$      & $r_{\rm cut}$  & $v_{\rm disp}$    \\
          & ($''$)            & ($''$)            &                       &  (deg)             &   (kpc)            &  (kpc)       & (km/s)      \\
\hline
DM halo    & $-0.6{\pm}1.6$    &  $2.1{\pm}1.4$    &  $0.18{\pm}0.02$      &  ${-}49{\pm}5$    &  [$75$]            &   [$1000$]    &  $935{\pm}30$  \\
CG1        & [$0.0$]           & [$0.0$]           &  [$0.07$]             &  [$169$]          &  [$0.2$]           &   [$42$]      &  [$230$]        \\
CG2        & [$2.4$]           & [$-1.2$]          &  [$0.11$]             &  [$202$]          &  [$0.2$]           &   [$38$]      &  [$220$]     \\
\hline\hline
\label{table:mass_model_params}
\end{tabular}
\vspace{-0.4cm}
\caption{{\small Note: Numbers in square brackets are fixed in the fit.  Position angles are clockwise from North.}
}
\end{center}
}
\end{table*}

\subsection{Spatially Resolved Spectroscopy}
\label{sec:resolved_spectra}
\subsubsection{Nebular emission}

By fitting a double Gaussian profile of fixed separation (corresponding
to the redshifted separation of the doublet) and intensity but variable
width, redshift and intensities of the doublet to the collapsed
spectrum, the [O{\sc ii}] emission yields a systemic redshift of
4.8757$\pm$0.0010 and and intrinsic width of $\sigma$=100$\pm$20$\kms$
(in the rest-frame of the galaxy).  We assume that the
[O{\sc ii}] maps the systemic redshift and all velocities are given
with respect to this value.  The integrated emission line flux of
5$\pm$1$\times$10$^{-20}$$\Wm2$ suggests a star-formation rate
(corrected for lensing amplification, but uncorrected for reddening) of
12$\pm$2$\Msolyr$ (assuming the calibration of L([O{\sc ii}]) to SFR
given in \citealt{Kennicutt98}).

In Figure~\ref{fig:sinf_1dspec} we show the intensity distribution and
collapsed spectrum of the [O{\sc ii}] emission.  Intensity and velocity
maps were derived by fitting the [O{\sc ii}] emission line doublet in
each 0.25$''$ pixel using the same technique described above.  We used
a $\chi^{2}$ minimisation procedure, taking into account the greater
noise of at the positions of the sky-lines.  In cases where the fit
failed to detect the line the region was increased to 2$\times$2 pixels
(0.5$''\times$0.5$''$).  Using a continuum fit, we required a minimum
signal-to-noise of 3 to detect the line, and when this criterion is
met, we fit the [O{\sc ii}] emission line doublet with a Gaussian
profile of fixed separation allowing the central wavelength and
normalisation to vary.  In Figure~\ref{fig:sourceall} we show the
intensity distribution and velocity structure of the [O{\sc ii}]
emission line flux.

It is clear that the brightest components in the rest-frame UV are also
the brightest in the [O{\sc ii}] emission.  In the source plane we find
that the eastern most component is marginally redshifted relative to the
western component. Relative to the combined spectrum, the eastern
region has $v=-20\pm20\kms$ (with a velocity dispersion of
$\sigma$=80$\pm$18$\kms$), while the western component is at 
+20$\pm$15$\kms$ (and a velocity dispersion of 85$\pm$12$\kms$). The
quoted uncertainty here is dominated by possible variations in the 
[O{\sc ii}] doublet line ratios, which was left free in these fits.

Since the [O{\sc ii}] velocity dispersion ($\sigma$) reflects the
dynamics of the gas in the galaxies' potential well we estimate the
dynamical mass of the galaxy.  Following the same prescription of
\citet{Erb06a}, we use the velocity dispersion and spatial extent of the
[O{\sc ii}] flux to infer a dynamical mass of $\sim1\times10^{10}\Msol$
within a radius of 2\,kpc.  Clearly this mass has large uncertainties
since the mass depends on the mass density profile, velocity anisotropy
and relative contributions to $\sigma$ from random motions or rotation
and possible differences between the total mass and that of the tracer
particles used to measure it.  Nevertheless, it is worth noting that
the mass we derive is a factor of $\sim5\times$ smaller than the median
dynamical mass of z$\sim$3 LBGs from \citet{Erb06a} measured in exactly
the same way.

\subsubsection{Ly$\alpha$}

In Fig~\ref{fig:sourceall} we show the reconstructed source-plane
Ly$\alpha$ morphology of the arc.  It is immediately apparent that in
the source-plane, the Ly$\alpha$ is more extended than the [O{\sc ii}],
or the continuum light, extending over 11.9$\times$2.4\,kpc (FWHM).
Figure~\ref{fig:vimos_1dspec} shows the collapsed spectrum of the arc
overlaid with the expected position of the Ly$\alpha$ emission line for
the systemic redshift (we also mark the expected position of the UV ISM
lines which lie in regions of the sky free from strong OH airglow
emission).  The integrated Ly$\alpha$ line flux is
3.7$\pm$0.5$\times$10$^{-15}$erg\,s$^{-1}$cm$^{-2}$, corresponding to a
star-formation rate of 5$\pm$2$\Msolyr$ (unlensed).  In comparison the
1500\AA\ continuum flux suggests a star-formation rate of
3$\pm$2$\Msolyr$ \citep{Kennicutt98}.

\begin{figure}
  \centerline{
    \psfig{file=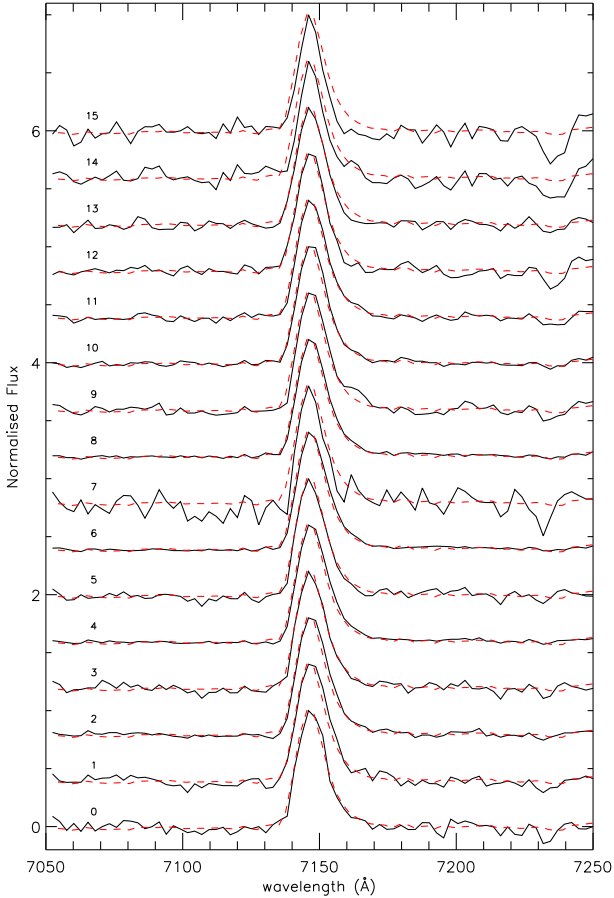,width=3.5in,angle=0}}
  \caption{One-dimensional spectra around the redshifted Ly$\alpha$
    emission from the $z$=4.88 galaxy from the fifteen regions shown in
    Figure~4 (black).  Each of the spectra has the composite spectra
    (scaled) and overlaid for comparison (red dashed).  The Ly$\alpha$
    emission in all of the spectra show an asymmetric profile, with the
    blue wing of the Ly$\alpha$ emission truncated making it appear
    redshifted by +200$\pm$40$\kms$ from the nebular ([O{\sc ii}])
    emission line.  Furthermore, the centroid of the Ly$\alpha$
    emission varies by less than 10$\kms$ (rest) across the whole
    galaxy image.
  }
\label{fig:vimos_multispec}
\end{figure}

From the collapsed spectrum of the galaxy we measure a redshift from
the centroid of the Ly$\alpha$ emission line and measure a velocity
offset from the systemic redshift of +200$\pm$40$\kms$.  Velocity
offsets of this magnitude between Ly$\alpha$ and nebular emission lines
(such as [O{\sc ii}] or H$\alpha$) are a common feature of galaxies at
$z\sim$3 \citep{Shapley03,Erb03}.  In both the collapsed spectrum
(Fig.~\ref{fig:vimos_1dspec}) and the spatially resolved spectra
(Fig.~\ref{fig:vimos_multispec}) the strong Ly$\alpha$ from the galaxy
shows an asymmetric profile in which the blue wing of the Ly$\alpha$ is
absorbed, presumably giving rise to the apparent redshift offset
between the [O{\sc ii}] and Ly$\alpha$.  It is also clear that there is
an extended wing of emission in the red side of the line.  

Asymmetry in the Ly$\alpha$ emission line spectra of LBGs at
$z\sim$2--3 are common and often give rise to velocity offsets of
several hundred km/s.  These velocity offsets are usually attributed to
galactic-scale outflows, presumably powered by supernovae.  However,
since the optical depth of the Ly$\alpha$ forest is an order of
magnitude larger at $z=5$ than at $z=3$, (where most spectroscopic
studies have taken place) we first investigate whether the IGM could
cause the absorption in the blue wing of the emission line.

At $z=5$ the mean optical depth of the Ly$\alpha$ forest is 1.7$\pm$0.2
\citep{Fan06} which corresponds to a transmitted flux of 20\%.  We
therefore investigate whether the Ly$\alpha$ profile could be explained
as the result of average IGM absorption, or whether the absorption
needed to be due to material associated with the lensed galaxy.  The
mean optical depth is measured by averaging the absorption of
individual clouds on scales of 6000$\kms$ (6\,Mpc comoving).  We fit
the Ly$\alpha$ emission line with an underlying Gaussian emission line
profile (centred on the systemic redshift of the galaxy as measured
from the [O{\sc ii}] emission), convolved with a step function which
models the average IGM (all profiles are then convolved with the
instrumental resolution).  We also include a broad Gaussian emission
line profile (also centred at the systemic redshift) to account for the
extended red wing of emission.  Allowing the redshift of the IGM
component to vary we find that an acceptable match emission line shape
is best matched when the IGM step function is placed between -80 and
+10$\kms$ from the galaxy (between -85\,kpc and +10\,kpc comoving for
out adopted cosmology). This is inconsistent with our assumption of a
large-scale average absorption, implying that the absorption we see
results from an individual cloud (or clouds) lying close to the target
galaxy.

Moreover, a simple step function fails to adequately reproduce the 
line shape of the blue wing. A better fit 
is obtained if the line shape is instead modified by a single absorbing cloud.
We model the emission profile by Gaussian profile combined with 
a Voigt profile absorber in the blue wing.  
In this fit the wavelength of the underlying best fit Gaussian 
profile (blue dotted line in Fig~\ref{fig:vimos_1dspec}) was allowed to 
vary, but the fitted centroid at 7143.0$\pm$0.5\AA\ ($z=$4.8758$\pm$0.0001)
is in excellent agreement with the nebular emission line redshift as measured
from [O{\sc ii}].  This underlying Gaussian profile also has a width of
$\sigma$=4.8\AA\ ($\sigma$=200$\kms$ in the rest-frame of the galaxy).
The absorber in the blue wing of the Ly$\alpha$ emission
(green dashed line) has a centroid of 7138.0$\pm$0.1\AA\ 
($z$=4.8717$\pm$0.0001, or $\Delta v=-500\kms$ relative to the
intrinsic emission) and a width of $\sim$100$\kms$.  Since the flux
in the Voigt profile reflects the column density of the neutral gas, we
infer $N_{\rm{HI}}=1.6^{+2.5}_{-1.1}\times 10^{19}\,$cm$^{-2}$ and
$\sigma$=110$\pm$50$\kms$.  The large uncertainties in the value of
$N_{\rm{HI}}$ and $\sigma$ reflect that fact that the inferred value of
$N_{\rm{HI}}$ lies between the flat and linear part of the curve of
growth for Hydrogen.

With this two component model we are able to fit the asymmetry in both
the collapsed and spatially resolved spectra.  However, even with two
components there remains excess flux in the red wing of the emission
line.  To compensate for this emission, we also fit a third Gaussian
profile component in the red wing, again allowing the redshift, width
and intensity to vary.  The improvement between the three fit and two
fit models is $\Delta\chi^{2}=12-38$ ($\sim3-6\sigma$; depending on
whether we fit the individual spectra or the combined spectrum).  The
best fit model therefore includes a third component with redshift
centroid at 7153.44 ($z$=4.8860) and a width of $\sigma$=260$\kms$.
The velocity offsets of the three components are summarised in Table~2.
As we discuss in \S~\ref{sec:discussion}, this third component may
naturally arise from Ly$\alpha$ photons back-scatter on the interior of
an outflowing shell.  Thus the best-fit model to the asymmetric
Ly$\alpha$ emission is one in which the Ly$\alpha$ photons are
generated in similar regions (and at the same redshift) as the [O{\sc
  ii}], but the emission profile is modified by foreground neutral
material.  This foreground material absorbs/scatters photons from the
blue wing of the line, causing an apparent redshift offset between the
nebular emission and Ly$\alpha$.

The description of the line profile given above is typical of the
emission profile seen in other Lyman break galaxies. The key advantage
of observing the lensed system is that we are able to spatially resolve
the emission line, and hence investigate the spatial variations in the
absorption and emission.  We divide the IFU data cube into the spatial
regions shown in Fig.~\ref{fig:sourceall}. For reference these have
been labelled 0-15.  The resulting spectra are shown in
Fig.~\ref{fig:vimos_multispec}.  The most striking feature of this
figure is that (apart from variations in intensity) the line profile
appears remarkably constant across the galaxy.  For each spatial
element, we measure the centroid of the emission and find that the
intrinsic emission line centroid varies by less than 10$\kms$
(rest-frame) across the whole galaxy image (this value was derived by
both fitting the 2 and 3 component models above).  Indeed, as
Fig.~\ref{fig:vimos_multispec} shows, the asymmetry in the Ly$\alpha$
profile is evident across the whole galaxy.  Applying a free fit, each
of the individual spectra is consistent with the mean, with the maximum
$\Delta\chi^2$=1.3 (and an average $\Delta\chi^2\sim$0.5).  From the
position of the Voigt absorber, we also note that the blue-edge of the
emission line cuts off at a wavelength corresponding to variations of
less than 20$\kms$ at a redshift of 4.8717$\pm$0.0001.  This constancy
in the line shape and centroid across the galaxy image is discussed
further below.

\subsubsection{UV absorption}

In the integrated rest-frame UV continuum spectrum of the galaxy, we
also detect the weak Si{\sc iv}($\lambda\lambda$1393,1402.77)
absorption features.  Whilst these are very weak in our spectra (the
S/N in continuum in regions free from strong sky emission is only
$\sim$3), the separation between the doublet is 9.80$\pm$0.15\AA\
(rest), is in excellent agreement with the expected value of 9.77.  To
calculate the redshift of these absorption lines we cross correlate the
spectrum around 8200\AA\ with the rest-frame UV composite spectrum of
Lyman break galaxies from \citet{Shapley03} and find that these
absorption lines are blue-shifted by $-$400$\pm$100$\kms$
($z$=4.870$\pm$0.0007) from the systemic redshift as defined by the
[O{\sc ii}] emission line (Figure~\ref{fig:vimos_1dspec}).  It is worth
noting that in the LBG composite spectrum from \cite{Shapley03} the
interstellar absorption features (O{\sc i}$\lambda$1303+Si{\sc
  ii}$\lambda$1260) are detected with similar significance as the
Si{\sc iv}.  However, at $z=4.88$ these are redshifted to 7656\AA\ and
into the middle of the Frauenhoffer A-band and therefore are therefore
not expected to be detected.

The observed velocity offsets between the nebular ([O{\sc ii}]),
Ly$\alpha$ and UV ISM lines are therefore consistent with the situation
seen in most high-redshift Lyman Break Galaxies where velocity offsets
of several hundred km/s have been measured
\citep[e.g.\,][]{Erb03,Steidel04}.  These velocity offsets are
attributed to starburst driven superwinds and we base much of our
interpretation on this model.

\begin{table}
\begin{center}
  {\footnotesize {\centerline{\sc Table 2.}}  {\centerline{\sc
        Properties of the emission and absorption lines}}
\begin{tabular}{lccccccc}
\hline
\hline
\noalign{\smallskip}
Component                    & $z$          & $\Delta$v            & $\sigma$   &  v$_{shear}$ \\
                             &              & $\kms$               & $\kms$     &   $\kms$   \\
\hline
[O{\sc ii}]                  & 4.8757[5]    & 0                    & 100$\pm$30 &  $-$30$<v<$30   \\
Ly$\alpha$ (Principle)       & 4.8760[5]    &  +20$\pm$50          & 230$\pm$40 &  $\lsim$20     \\
Ly$\alpha$ (red wing)        & 4.8860[15]   & +520$\pm$150         & 260$\pm$80 &  $\lsim$50     \\
Ly$\alpha$ (blue abs. )      & 4.8680[10]   & $-$500$\pm$100       & 110$\pm$30 &  $\lsim$20     \\
Si{\sc iv} absorption        & 4.8700[10]   & $-$400$\pm$100       & $\sim$480  &  $\lsim$150       \\
\hline\hline
\label{table:collapsed_spec_properties}
\end{tabular}
\caption{Note the value given in the [] $z$ column is the error in the
last decimal place. v$_{shear}$ denotes the maximum velocity
gradient/shear in the given emission/absorption line.
}
}
\end{center}
\end{table}

\section{Discussion}
\label{sec:discussion}

Before attempting to interpret the observations, we briefly review the
results of the observations.  The [O{\sc ii}] emission maps the
star-forming regions, which appear to have an elongated (bar-like)
morphology with a spatial extent of only 2.0$\times$0.8\,kpc, which
appears to contain all of the stars in the galaxy.  We use the [O{\sc
  ii}] emission to infer a systemic redshift of 4.8757$\pm$0.0005.
Using the [O{\sc ii}] emission we infer an integrated star-formation
rate of 12$\pm$2$\Msolyr$ \citep{Kennicutt98} and the spatial extent
and line width of the [O{\sc ii}] suggest a dynamical mass of
$\sim$1$\times$10$^{10}\Msol$ within 2\,kpc.  Turning to the rest-frame
UV spectrum: the Ly$\alpha$ emission is much more extended than the
[O{\sc ii}], extending over 11.9$\times$2.4\,kpc (i.e. at least a
factor of 4$\times$ the spatial extent of the [O{\sc ii}]).  Moreover,
the peak of the Ly$\alpha$ emission is redshifted by +200$\pm$40$\kms$
from the [O{\sc ii}].  Across the whole spatial extent, the Ly$\alpha$
emission line centroid varies by less than 10$\kms$ (rest-frame) and
the line profile is remarkably constant over the same spatial extent.
The Ly$\alpha$ emission (both in the collapsed spectrum and spatially
resolved) has an asymmetric profile which is best explained by a three
component model.  The underlying Gaussian emission line profile has a
centroid in excellent agreement with the nebular emission, an absorber
which is blue-shifted by $-$500$\pm$100$\kms$ from the systemic
redshift and an extended ``red wing'' of emission is redshifted by
+520$\pm$150$\kms$.  Finally, the UV ISM lines of Si{\sc iv} are
blue-shifted by -400$\pm150\kms$ from the systemic redshift, although
it has not been possible to derive any spatial information across the
galaxy from these weak lines.

\subsection{A starburst with a bi-conical outflow?}

The observed velocity offsets between the [O{\sc ii}], Ly$\alpha$ and
UV ISM lines in RCS0224arc are consistent with situation seen in most
high redshift Lyman Break Galaxies in which the presence of
``superwinds'' (driven by the collective effects of star-formation and
supernovae) naturally explain the observations.  Our spatially resolved
spectra therefore offer us the opportunity to study the spatial
structure and energetics of the outflowing material in this galaxy.

The velocity offset between the [O{\sc ii}] and transmitted Ly$\alpha$
emission profile are usually interpreted in terms of a starburst driven
outflow. This model provides a good description of the integrated line
emission profiles of high redshift galaxies and more detailed
observations of local star bursts (eg.,
\citealt{Pettini02,Shapley03,Tenorio-Tagle99,Heckman00,Grimes06}).
Here we investigate the implication for this model from the spatially
resolved profile that we have observed. We base our interpretation on
the specific models of \cite{Mas-Hesse03} which assume that the
spatially compact [O{\sc ii}] nebular emission traces the star forming
regions in the nucleus of a star-bursting galaxy.

In these models, the velocity offsets between [O{\sc ii}] and
Ly$\alpha$ is generated by a star forming region embedded in an
outflowing bubble. The models have been extensively investigated in
local starburst galaxies where similar spatially resolved data is
available.  Indeed, there is strong similarity between the data we
present, and the local galaxy Haro 2 studied by \citet{Mas-Hesse03}.
The spatial extend (500\,pc) and velocity shift ($v_{\rm
  shell}\sim200\,\kms$) seen in Haro 2 are smaller than the $z=4.88$
galaxy but otherwise the systems are remarkably similar.  Both show
Ly$\alpha$ emission that extends beyond the nebular emission region and
both show a blue absorption cut-off to the Ly$\alpha$ emission line and
a similar red-wing.  Crucially, both data sets show no detectable
variation in the shape of the Ly$\alpha$ emission profile across the
emission region.  It is remarkable that this galaxy, seen 1\,Gyr after
the Big-Bang is so similar to the local system at $z=0.0049$.

The basis of the interpretation is that the star-burst generates a
large over pressurised region which surrounds the nebula. Within the
ISM, the flow is Reighly-Taylor unstable so that the initial shell
breaks up leaving behind a denser gas clumps that may be responsible
for the extended Ly$\alpha$ emission region. The outflow tends to break
out of the galaxy along the axis of least resistance. In the case of a
disk (or flattened) galaxy this generates an approximately bi-conical
outflow.  When the outflow reaches the smoother, lower density gas
distribution in the halo of the galaxy, a new shell develops as the
outflow sweeps up material from the galaxy's halo, creating a screen
between the observer and the source. As the shell travels away from the
galaxy, it becomes dense enough to recombine and its Ly$\alpha$ opacity
becomes large.  \citet{Mas-Hesse03} interpret Haro~2 as being an
example of a system in these later stages, and the $z=5$ arc fits into
the paradigm well.

The model interprets the spatially integrated properties of the
emission and absorption lines seen in Figure~\ref{fig:vimos_multispec}
as follows: the main peak of Ly$\alpha$ emission comes from photons
emitted from the star forming regions. To reach the observer these must
pass through part of the foreground (blue shifted) shell. This
absorption causes the peak emission wavelength to appear redshifted
relative to the nebular emission lines.  In addition, photons may be
scattered or emitted from the receding shell.  Photons that are either
created on the inner surface of the shell (e.g. by UV irradiation from
the starburst), or multiply scattered within the receding shell so that
they acquire the mean velocity of the shell (see the discussion of
\citealt{Hansen06}) will be seen as redshifted by the observer.

However, while the integrated spectrum is certainly consistent with the
presence of an outflowing bubble, it gives no indication of its
physical size, or whether the bubble geometry is appropriate.  For
example, the integrated spectra could be equally well explained by a
large scale outflow, or a small bubble only just surrounding the
starburst region.  To investigate the energy of the outflow, and its
ability to escape the host galaxy's gravitational potential, we must
use spatial information in our spectra.

\begin{figure*}
  \centerline{
    \psfig{file=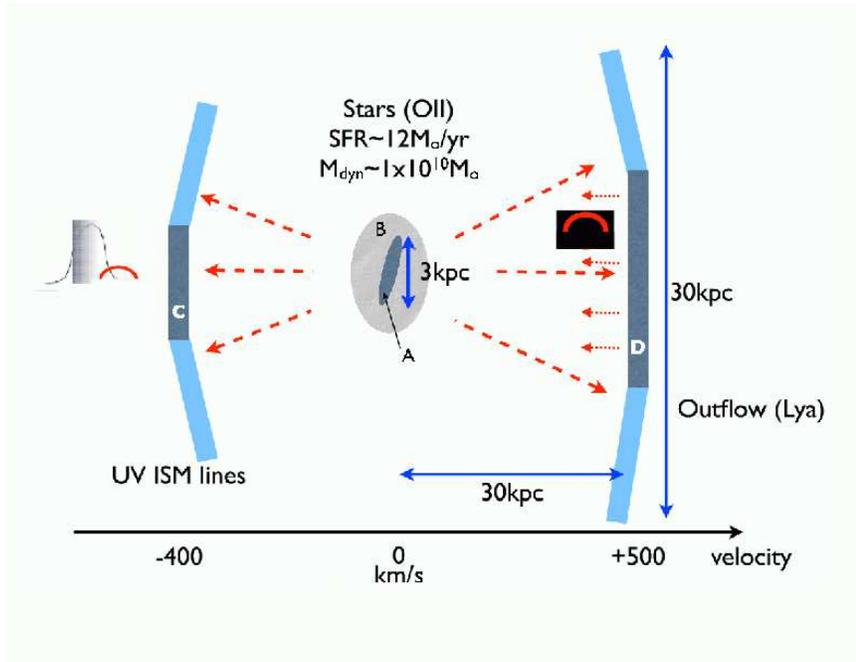,width=4.5in,angle=270}}
\caption{This figure illustrates the details of the proposed 
  configuration to explain the spatially resolved emission spectrum.
  The Ly$\alpha$ emission comes predominantly from photons generated
  in the star forming region `A'.  The Ly$\alpha$ emission region is
  physically larger than the due to recombination in dense gas clumps
  that have not yet been swept up and ionised by the outflow, and the
  resonant Ly$\alpha$ may not escape readily from the central regions
  of the starburst.  The outflow cavity 'B' contains predominantly
  ionised material flowing out of the galaxy.  It is hot and diffuse
  and therefore not seen directly.  However, the shell 'C' (travelling
  towards the observer) that has formed as the wind has swept up the
  interior of the cavity can be seen in absorption at $-v_{\rm shell}$
  against the central Ly$\alpha$ emission region.  The geometry of the
  outflow, and its extent compared the source mean that the projected
  velocity is almost constant across the illuminated portion of the
  shell.  Photons that scatter of the receding shell ('D'), or are
  produced by ionisation of its inner edge by UV photons from the
  starburst, are seen by the observer as a weak emission line
  (broadened by radiative transfer effects) at $+v_{\rm shell}$.  The
  bi-conical nature of the outflow is key to explaining why the
  redshifted emission is only seen over the portion of the shell
  directly behind region 'A'.  Redshifted photons from other regions of
  the receding shell must pass through the area of the galaxy that has
  not yet been ionised.  Even though they are redshifted relative to
  the galaxy, they will be absorbed since the absorbing column is
  likely to be extremely high.  Note that photons that pass through the
  central ionised region of the galaxy are unlikely to be absorbed by
  the foreground shell since they are redshifted by $+2v_{\rm shell}$.
}
\label{fig:cartoon}
\end{figure*}

We begin by considering the dominant contribution to the Ly$\alpha$
line from photons travelling directly towards the observer. The
wavelength of the blue cut-off in the line is determined by the outflow
velocity seen by the observer.  If the bubble is spherical and close to
the galaxy we would expect to see significant variation in the blue
cut-off across the emission region since the line of sight velocity of
the outflow depends on $\sqrt{1 - (b/R_s)^2}$, where $b$ is the impact
parameter and $R_s$ is the shell radius. Clearly the consistency of the
blue cut-off in Ly$\alpha$ across the galaxy implies that the shell is
much larger than the galaxy so that the illuminated portion of the
absorbing screen has a planar geometry.  We illustrate the likely
viewing geometry in Fig.~\ref{fig:cartoon} (cf., Fig.\ 18 of
\citealt{Mas-Hesse03}) for an outflow in the shell recombination phase
(phase 4 in their classification scheme).  In Haro~2, \citet{Legrand97}
are able to estimate the size of the shell from its low surface
brightness H$\alpha$ emission. They estimate that the projected
diameter of the shell is 2.5\,kpc, around five times the diameter of
the emission region. A similar ratio would seem appropriate for the
arc, although the shell could obviously be larger.

Our data also require an additional contribution from photons scattered
or created in the receding shell. For a large spherical shell, we would
expect to see a widespread low surface brightness emission at this
wavelength (e.g. Fig.~\ref{fig:cartoon}).  However, no such distributed
emission is seen in our data, or in Mass-Hesse et al.'s observations of
local systems.  Instead, by collapsing the datacube over a region
dominated by the excess flux in the red wing, we find that the
redshifted wing seems to have a similar angular distribution to that of
the Ly$\alpha$ emission coming directly from the galaxy.  However, we
can extend the model of Mass-Hesse et al. to explain the limited
spatial extent of this emission as a result of the bi-conical outflow
geometry.  Redshifted photons from most of the receding shell must pass
through the area of the galaxy that has not yet been ionised by the
starburst. Even though they are redshifted by $+v_{\rm shell}$ relative
to the galaxy, they will still be absorbed since the absorbing column
is likely to be extremely high.  Only in the ionised region surrounding
the starburst will the Ly$\alpha$ opacity be sufficiently low to allow
photons to pass though the otherwise neutral disk (photons that pass
through the ionised region are unlikely to be absorbed by the
foreground shell since they are redshifted by $+2v_{\rm shell}$). The
column density of the residual gas disk is also likely to be much
larger than that of the expanding shell.  The result of this geometry
is that the redshift wing will have similar spatial extent to the main
Ly$\alpha$ line.

The proposed geometry explains the spatial distribution of the spectra
without needing to invoke a special viewing angle. The only constraint
is that the system is viewed down through the outflow cone.  As the
inclination between the outflow axis and the viewer becomes
significant, we might expected to detect some rotation from the galaxy
(a small shear is suggested by our SINFONI spectra of the [O{\sc ii}]
emission), however, a shift of a several $10\,\kms$ in the {\it
  intrinsic} profile of the Ly$\alpha$ line is unlikely to be
observable since the line shape is so strongly shaped by the
blue-shifted absorption (which does not share the velocity shear of the
galaxy).  If the system is viewed at a larger angle, Ly$\alpha$ is
likely to be completely absorbed since the expansion velocity
perpendicular to the cone is small. In this case, it is unlikely that a
redshift would have been obtained for this arc.  It is also worth
noting that the outflow spends much of its time in this post-blow-out
phase. In earlier phases, the foreground shell will have lower column
density or may even show emission at its leading edge; in later phases,
as the shell slows down and stalls, Ly$\alpha$ does not escape at all.
This model does not require us to be observing the galaxy at a
particularly special phase of its evolution or at a special viewing
angle.

Whilst the proposed bi-polar outflow fits the observables, there are
other possible scenarios which would explain the velocity offsets we
measure: if the outflow were instead within the galaxy, and the
Ly$\alpha$ emission was non-Gaussian (e.g. a Lorentzian) then the red
wing of emission seen at $z=$4.8860 could be explained without a
receding shell.  In this model, the blue-shifted material could arise
due to an ISM outflow which is in the process of breaking out of the
galaxy.  However, the constancy of the Ly$\alpha$ emission line shape
and centroid is difficult to explain -- any velocity motion within the
star-forming regions should be reflected by the Ly$\alpha$, which is
not seen in our data.

\subsubsection{Energetics of the outflow} 

Although there are uncertainties in the geometry and column density of
the outflowing shell, it is nevertheless interesting to consider the
energetics of the outflow, in order to see if the bi-polar outflow
scenario is reasonable.  For a normal stellar initial mass function,
supernovae provide $\sim$10$^{49}$\,erg per solar mass of stars (eg.,
\citealt{Benson03}), thus for a galaxy with a star-formation rate of
12$\pm$2$\Msolyr$, $\sim$10$^{50}$\,erg are available per year and the
maximum extent of the Ly$\alpha$ emission is $\sim12$\,kpc diameter. We
estimate the distance of the shell from the galaxy by assuming that the
linear size of the region covered by the swept up shell is at least $q$
times the size of the maximum extent of Ly$\alpha$ emission region.
For Haro~2, the shell is at least three times the spatial extent of the
star-forming regions.  If we adopt $q<3$ then we would expect to detect
variations in the velocity of foreground absorber. This suggests a
linear size for the end of the cone of $\gsim 30$\,kpc.  Adopting an
opening angle for the outflow of $\sim60^{\circ}$, we estimate that the
shell must be located $\gsim30$\,kpc from the starburst region.

At a velocity of $\sim500\,\kms$ it takes 60\,Myr to travel out to a
distance of 30\,kpc; within this time, supernovae will provide a total
energy of $\sim$6$\times$10$^{57}$\,erg.  Clearly there are a number of
uncertainties in this estimate: not least that the shell may have
decelerated from an initial higher velocity which would reduce the
timescale.  Nevertheless, we can estimate the mass of the outflow via:
$M_{outflow} = N_{HI}\times M_{H}/x$ where $N_{HI}$ is the column
density, $A$ is the area of the cone and $M_{H}$ is the atomic mass of
Hydrogen.  Adopting the value of the observed column density we derived
in \S\ref{sec:resolved_spectra} of $N_{\rm{HI}}=1.6 \times
10^{19}\,$cm$^{-2}$ and that the cone is uniform over an area of
700~kpc$^{2}$, the total mass of the outflow is $1.8 \times 10^{8} /x
\Msol$, where $x$ is its H{\sc i} fraction.  The kinetic energy of the
outflow is then $E=\frac{1}{2}mv^{2} = 5 \times 10^{56}/x$\,erg.  Thus,
the outflow is energetically feasible if the neutral fraction is
greater than 10\%, or the column density of the shell is lower than we
have assumed. (Note, however, that these two factors well tend to play
off against each other: a lower column density will tend to have a
lower neutral fraction).

It is interesting to note that the implied mass outflow rate is
therefore $\sim3/x\Msolyr$.  If we assume $x$=0.1, then the mass
outflow rate is very efficient at ejecting baryons from the galaxy,
with the mass loading of the wind being more than twice the
star-formation rate of the galaxy.

The estimates above are clearly uncertain. Nevertheless, it is
interesting to see how far the expanding material might be expected to
travel after the end of the starburst.  We assume a halo circular
velocity of 150\,$\kms$ and a wind speed of 500\,$\kms$ located 30\,kpc
from the center of the halo. If the wind flows freely out of the galaxy
it is completely unbound.  However, the wind will slow down
dramatically if it sweeps up material.  To make an estimate of how far
the wind will travel we assume that the wind expands into an NFW halo
with a baryon fraction of 18\% and that the wind has an opening angle
of 60$^{\circ}$. Even though the wind rapidly gains mass as it expands,
it reaches a minimum of 160\,kpc (almost 1 co-moving Mpc) before
stalling.  Thus, this starburst driven wind has enough energy to
pollute a volume of approximately $\sim$3Mpc$^{3}$.

\section{Conclusions}

The issues of metal ejection and feedback are some of the most
outstanding problems in galaxy formation.  One of the most important
recent observational breakthroughs is that most high-redshift
proto-galaxies appear to be surrounded by ``superwinds'' which are
expelling material from the galaxy disc.  However, at high redshift
(where star formation efficiency was at its peak and therefore this
phenomenon is likely at its peak activity), observational evidence is
usually based on measuring the velocity offsets between the nebular
emission (such as [O{\sc ii}] or H$\alpha$) and rest-frame UV lines
such as Ly$\alpha$ or UV ISM lines. These observations are usually
based on longslit observations which lack spatial information, which is
crucial if we are to understand the dynamics and fate of the outflowing
gas.

In this paper, we have spatially resolved and mapped the dynamical
properties of a highly magnified gravitationally lensed galaxy at
$z$=4.88.  The main features are described below:

$\bullet$ The lensing amplification is a factor 16$\pm$2
($\Delta$m=3.0$\pm$0.2 magnitudes).  In the (reconstructed)
source-frame the {\it HST} imaging suggests the galaxy is only
$\sim2\times0.8$\,kpc FWHM.  Moreover, the lensing corrected $I$-band
magnitude is $I$=25.2.  For comparison, an $L^{*}$ Lyman-break galaxy
at $z\sim$ has $i$=25.4, suggesting this galaxy is typical of galaxies
at these early times.  Due to the lensing amplification we are able to
study a typical $z\sim$5 galaxy at high signal-to-noise and spatially
resolve the galaxy on $\lsim$200pc scales.

$\bullet$ The [O{\sc ii}]$\lambda\lambda$3726.2,3728.9 emission line
maps the (systemic) nebular emission and shows a peak-to-peak velocity
gradient of $\lsim$60$\kms$ across 3\,kpc in projection.  The
integrated [O{\sc ii}] emission line flux suggests an integrated
star-formation rate of 12$\pm$2$\Msolyr$.  The velocity shear, spatial
extent and velocity dispersion of this nebular emission suggest a
dynamical mass of $\sim1\times10^{10}\Msol$ within a radius of 2\,kpc.

$\bullet$ The Ly$\alpha$ emission is much more extended than either the
UV continuum or [O{\sc ii}] emission, extending over
11.9$\times$2.4\,kpc (FWHM).  The Ly$\alpha$ emission has a asymmetric
profile causing it to appear redshifted from the systemic velocity by
$+$200$\pm$40$\kms$.  Across the whole galaxy image both the shape and
centroid of the Ly$\alpha$ emission are remarkably constant with the
centroid of Ly$\alpha$ varying by $<$10$\kms$.

$\bullet$ Both the collapsed and spatially resolved Ly$\alpha$ emission
are best fit with a 3 component model: (i) an underlying (Gaussian)
emission profile which has a redshift and width in excellent agreement
with the nebular emission; (ii) an absorber with a Voigt profile at
$-$500$\kms$ and a column density of 1.6$\times$10$^{19}$cm$^{2}$,  and
(iii) a broad redshifted emission line component at +400$\kms$.

$\bullet$ In the collapsed rest-frame UV spectrum, the weak UV ISM
lines of Si{\sc iv} are blue-shifted from the systemic velocity by
$-$400$\pm$100$\kms$ (the velocity of these lines are in agreement with
the velocity of the absorber seen against Ly$\alpha$).

The velocity offsets between the nebular emission and spatially
extended Ly$\alpha$ are comparable to those seen in most high redshift
LBGs at similar redshifts \citep{Shapley03,Erb03} and are usually
attributed to galactic-scale outflows.  Hence, this leads us to
interpret the results in the context of an evolving starburst outflow
and we illustrate the likely viewing geometry in
Fig.~\ref{fig:cartoon}.  In this model, the [O{\sc ii}] emission is
assumed to trace the underlying (star-forming) region embedded in an
bi-conical outflowing bubble in which the starburst generates an
over-pressurised region within the ISM which becomes Rayleigh-Taylor
unstable in which the initial shell breaks up, leaving behind denser
clumps which may be responsible for the extended Ly$\alpha$ region.
Any outflow will break out of the galaxy along the axis of least
resistance and hence for gas in a planar or bar-like galaxy this
generates an approximately bi-conical outflow.

The proposed geometry naturally explains the spatial distribution of
the spectra and the velocity offsets and structures we observe.  We
consider the energetics of the outflow and show that the proposed model
is self-consistent and energetically feasible. The energy of the wind
is such that it plausibly reaches $\sim$160\,kpc (almost 1 co-moving
Mpc) before stalling. At this point, it will have polluted a comoving
volume of nearly $3\,$Mpc$^{3}$ with metal enriched material.

Whilst these observations are based on a single galaxy, our results
clearly show the power of combining gravitational lensing with optical
and near-infrared observations to probe the star-formation activity,
masses and feedback processes in typical high redshift galaxies in
great detail.  The next step is to generate a statistically useful
sample to gauge the prevalence outflows from these young galaxies.
In turn these measurements will help us understand the wide-spread
enrichment of the early universe, and may explain why only 10\% of baryons 
cool to form stars.

\section*{acknowledgments}

We would like to thank the anonymous referee for their suggestion which
significantly improved the content and clarity of the paper.  We also
thank Alice Shapley and collaborators for allowing us to use their
rest-frame UV composite spectrum of LBGs, Markus Kissler-Patig for
advice and support for the ESO/IFU observations and Tom Theuns, Howard
Yee, Tracy Webb and Erica Ellingson for useful discussions.  The VIMOS
and SINFONI data are based on observations made with the ESO Telescopes
at the Paranal Observatories under programmes 074.A-0035 and
075.B-0636.  AMS acknowledges support from a PPARC fellowship, RGB
acknowledges a PPARC Senior Fellowship, GPS and IRS acknowledges
support from Royal Society University Research Fellowships and JPK
thanks support from CNRS.

\bibliographystyle{apj} 
\bibliography{/Users/markswinbank/Projects/ref}

\appendix
\section{Serendipitous Background Galaxies}
\label{sec:serendip}

Since we are able to simultaneously survey all of the critical lines
from $z$=0 to $z$=7 with the VIMOS IFU and part of the $z$=1.5 to
$z$=14 critical lines with the SINFONI IFU we exploit the SINFONI and
VIMOS data to search for serendipitous sources behind the lensing
cluster.  We identify seven candidates between $z=$1.5 and $z=$5.5 from
both optical and near-infrared spectroscopy, at least three of which
have either optical or near-infrared broad-band counterparts.  One of
these sources is a radial counter-image of the $z$=4.88 arc which is
then used to provide strong constraints on the lens modelling
(\S~\ref{sec:mass_modelling}).  The remaining sources in the VIMOS
datacube are identified as [O{\sc ii}] emission at $z$=0.99 and
Ly$\alpha$ at $z$=3.66.  The SINFONI IFU emitters are tentatively
identified as either H$\alpha$ at $z$=1.5 to $z$=2.7 or [O{\sc ii}] at
$z\sim$5.3 (Table~\ref{table:serendip_sources}).  These observations
show the power which integral field spectroscopy will have in the
future in finding high-redshift galaxies behind cluster cores (e.g.
\citealt{Ellis01,Kneib04b}).  The observations presented here also
allow an insight of the deeper observations that will be made possible
with even larger area integral field spectrograph's on the VLT in the
near future (e.g.\, MUSE; \citealt{Henault04}) or the mosaic mode of
KMOS \citep{Sharples04}).  These observations will open the doorway to
probing (and spatially resolving) the properties of the primeval
galaxies responsible for reionization around $z$=8-12.

\subsection{Identifying Serendipitous Galaxies in the datacube}

Using the IFU coverage of the cluster cores, we attempt to identify
serendipitous sources in the IFU datacubes.  For the wavelength range
of VIMOS (5000\AA-1$\mu$m), we are able to search for Ly$\alpha$
between $z$=3.11 and $z$=6.8 and [O{\sc ii}] between $z$=0.34 and
$z$=1.57.  The wavelength coverage of the SINFONI IFU observations
($\lambda$=1.451--2.463$\mu$m) also allow us to search for [O{\sc ii}]
emission between redshifts $z=$2.90 to $z=$5.61, H$\alpha$ between
$z=$1.21 and $z=$2.75 and even Ly$\alpha$ between $z=$10.9 and
$z=$19.3).

\begin{figure}
  \centerline{
    \psfig{file=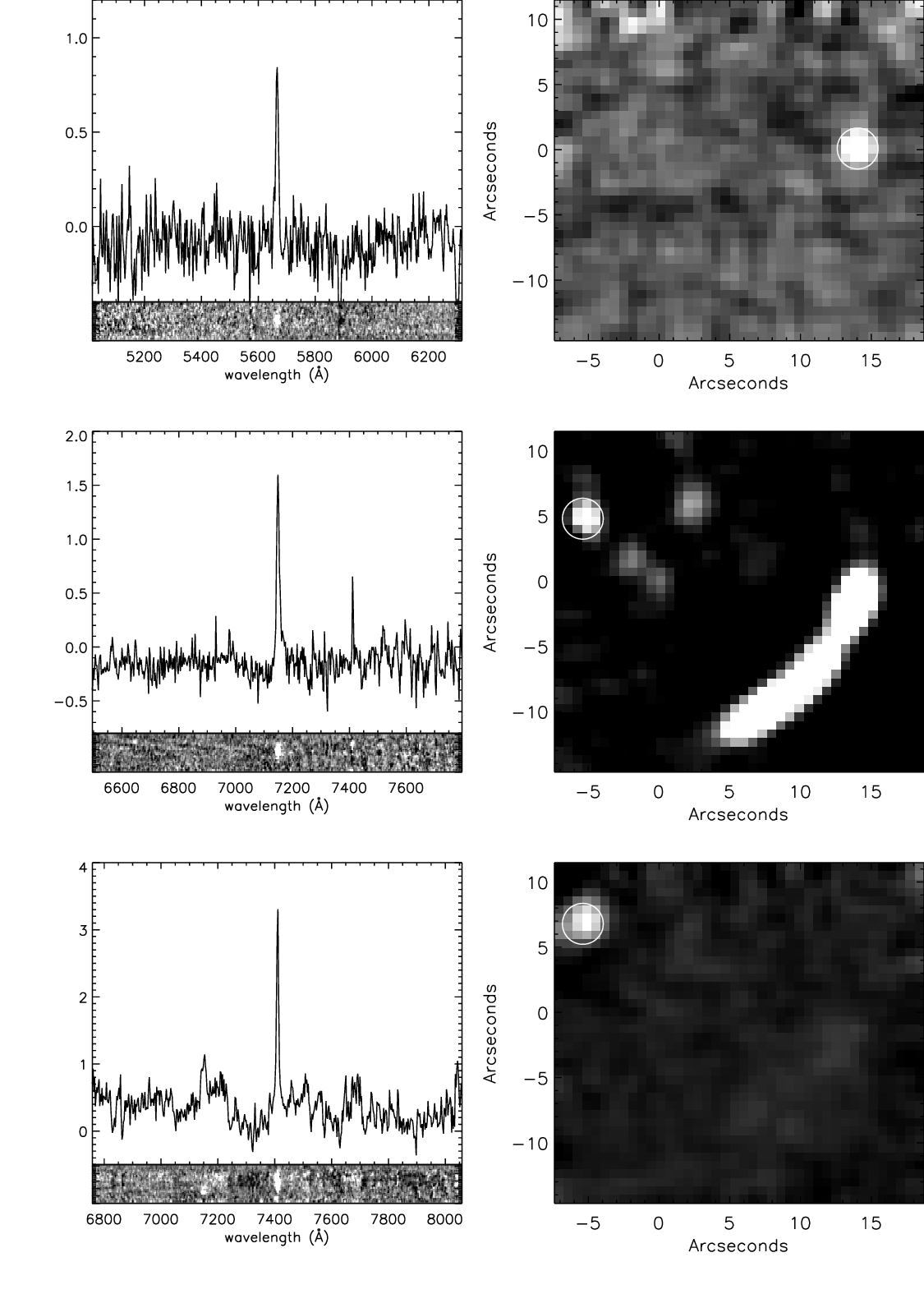,width=3.0in,angle=0}}
  \caption{Spectra of the three background, serendipitous sources in
    the VIMOS IFU field of view.  In all three panels we show the
    collapsed, one-dimensional spectrum, the two-dimensional spectrum
    (generated by unwrapping the datacube around the object) as well as
    the location of the object in the datacube (right).  {\it Top:} The
    super-imposed foreground galaxy (labelled A1 in Fig.~1) on the
    third image of the $z$=4.88 arc has strong line emission at
    5600\AA\ which we identify as Ly$\alpha$ yielding a redshift
    $z$=3.66; {\it Middle:} The $z$=4.789 arc is a radial counter-image
    (labelled R1 in Fig.~1) of the arc and is used as a primary
    constraint on the lens model in \S~\ref{sec:mass_modelling}. {\it
      Bottom:} The blue-disk galaxy (labelled A2 in Fig.~1) is lensed
    source at $z$=0.99 from identification of [O{\sc ii}] at
    7400\AA.  }
\label{fig:vimos_serendip}
\end{figure}

\begin{figure}
  \centerline{
    \psfig{file=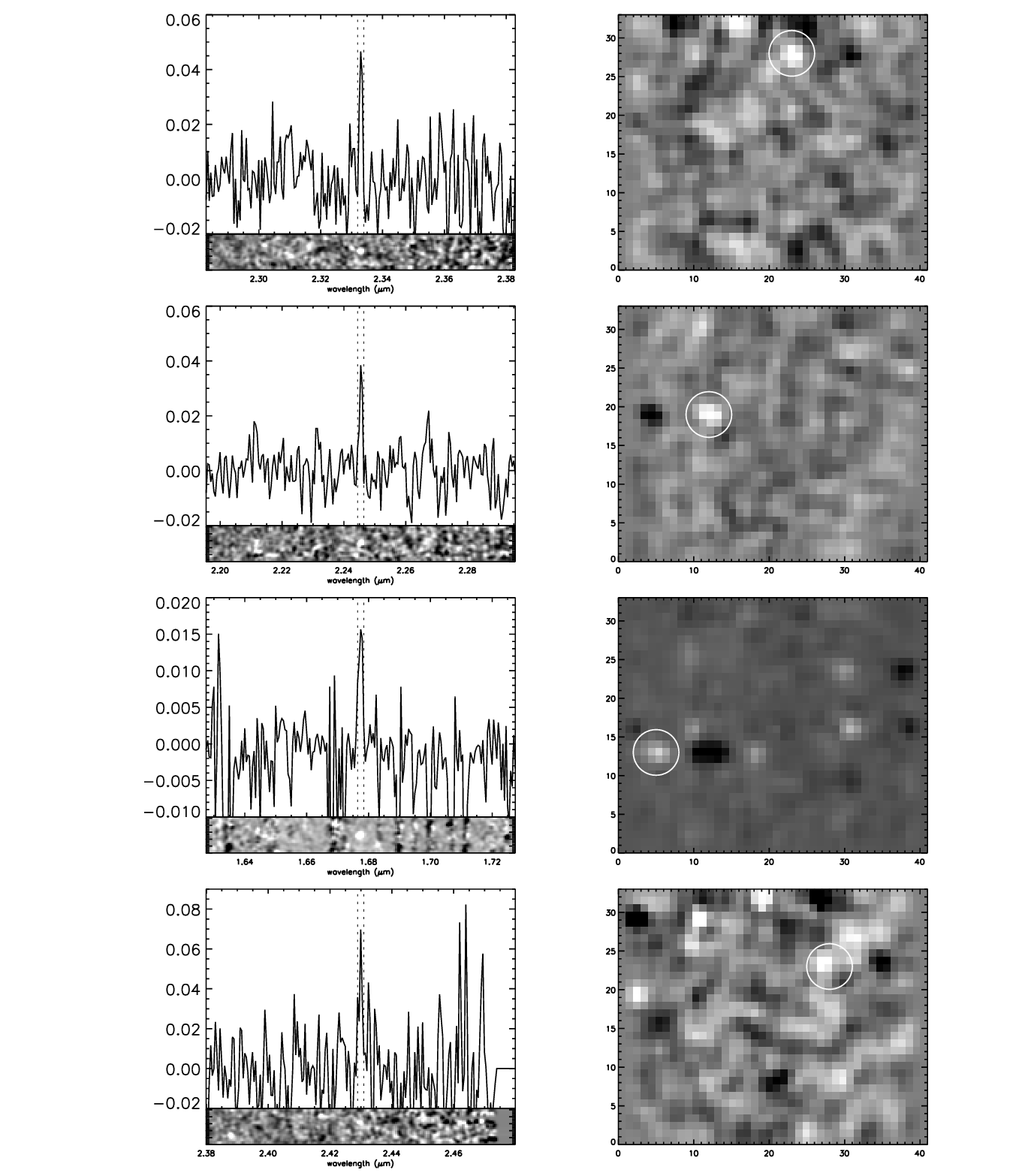,width=3.5in,angle=0}}
  \caption{Spectra of the four background, serendipitous sources in
    the SINFONI IFU field of view.  In the left hand panels we show the
    one-dimensional spectra with the two-dimensional spectra below.  In
    the right hand panel we show the two-dimensional (narrow-band)
    images which show the positions in the datacubes for these sources.
    The positions of these sources are also shown in Fig.~1 and details
    are also given in Table~\ref{table:serendip_sources}.  The images
    in the right hand panel are generated by collapsing the datacubes
    over the wavelength range shown by the two dashed lines in the
    one-dimensional spectra.}
\label{fig:sinf_serendip}
\end{figure}

To identify serendipitous line emission we scan each pixel of the
datacube.  At each wavelength in each spectrum we calculate the noise
over $\pm$150\AA\ and demand a signal-to-noise of four to detect an
emission line in at least three of the eight adjacent pixels (we mask
out any regions of strong sky emission from our analysis completely).
If line emission is detected, we then fit the emission line with a
single Gaussian profile and require that the emission line must have a
line width greater than that of the sky at the given wavelength.  Using
this technique, we identify three candidates in the VIMOS IFU datacube
(excluding [O{ \sc ii}], [O{\sc iii}] or H$\beta$ emission from the
cluster members) and four candidates in the near-infrared (SINFONI)
datacube.  We show the one and two-dimensional spectra (as well as
their location in the datacube) in
Fig.~\ref{fig:vimos_serendip}\&~\ref{fig:sinf_serendip} and briefly
review their properties and likely redshifts here.

\noindent
\smallskip
{\bf Optical}\\
\noindent
{\it VIMOS Source 1 (R1):} The first serendipitous source is the radial
counter-image of the $z$=4.879 arc (labelled R1 in
Fig.~\ref{fig:hst+vimos_col}).  The position of this counter-image is
used as an extra constraint in the lens modelling in
\S~\ref{sec:mass_modelling}.  \smallskip

\noindent
{\it VIMOS Source 2 (A1):} In the {\it HST} imaging of the cluster, the
third counter-image of the $z$=4.789 arc (at approximately
+6$''$,+14$''$) shows a bright knot superimposed onto the arc which
does not appear in any of the other counter-images of the arc and
therefore (as noted by \citealt{Gladders02}) is most likely a
foreground galaxy super-imposed onto the arc. This galaxy has a $V$-band
magnitude of $V_{606}$=25.6$\pm$0.2
and we identify an emission line at 5665\AA\ with an emission line flux
of 1.4$\pm$0.5$\times$10$^{-16}$erg\,s$^{-1}$\,cm$^{-2}$ and an
observed equivalent width of 700$\pm$150\AA.  If the line is identified as
[O{\sc ii}], then a redshift of 0.51 yields an EW of $\sim$470\AA, which
is high compared to local star-forming galaxies.  We therefore suggest
that the emission is Ly$\alpha$ at a redshift of $z$=3.66 with a
rest-frame EW of 150$\pm$50\AA.  The emission line width (FWHM=12.5\AA)
suggests a line width of 220$\pm$60$\kms$.  \smallskip

\noindent {\it VIMOS Source 3 (A2):} The third background galaxy
detected is the blue disk-like galaxy labelled A2 in
Fig.~\ref{fig:hst+vimos_col}.  We detect strong emission at 7409\AA\,
but no other significant ($>$5$\sigma$) emission lines in the
one-dimensional spectrum.  The emission line also has a velocity shear
of $\sim$5\AA\ across the $\sim$1.5$''$.  If the emission is identified
as [O{\sc ii}]$\lambda$3727 a redshift of 0.9855 is derived, placing
the galaxy behind the cluster, and probably lensed.  However, it is
also possible that the emission is H$\alpha$ at $z=$0.129, although we
place limits on the emission line ratios of H$\alpha$/H$\beta$$>$12 and
H$\alpha$/[O{\sc iii}]$\lambda$5007$>$6.  Identification as [O{\sc
  iii}]$\lambda$5007 is ruled out due to the non-detection of [O{\sc
  iii}]$\lambda$4959.  Alternatively, if the emission is H$\beta$ at
$z=$0.524, we constrain the emission line flux ratios as
H$\beta$/[O{\sc iii}]$\lambda$5007$>$8 and H$\beta$/[O{\sc
  ii}]$\lambda3727$$>$17.  We therefore suggest the most likely
identification is [O{\sc ii}]$\lambda$3727 at $z=$0.9855.  \smallskip

\begin{table*}
\begin{center}
{\footnotesize
{\centerline{\sc Table A1.}}
{\centerline{\sc Properties of the Serendipitous Sources}}
\begin{tabular}{lccccccc}
\hline
\hline
\noalign{\smallskip}
ID          & RA          & Dec          & $\lambda_{obs}$ & Line Flux              & Likely redshift & EW$_{observed}$ \\
            &             &              & ($\mu$m)        & (x10$^{-16}$erg\,s$^{-1}$)   & & (\AA) \\
\hline
\hspace{-0.3cm} VIMOS \\
A1         & 02:24:33.862 & -00:02:17.61 & 0.5665          & 1.4$\pm$0.5     & 3.660 (Ly$\alpha$)  & 470$\pm$60 \\
A2         & 02:24:34.602 & -00:02:34.80 & 0.7400          & 6.0$\pm$1.0     & 0.986 ([O{\sc ii}]) & 102$\pm$20 \\
R1         & 02:24:34.466 & -00:02:34.77 & 0.7148          & 4.5$\pm$0.8     & 4.879 (Ly$\alpha$)  & 720$\pm$30 \\
\hline
\hspace{-0.3cm} SINFONI \\
1          & 02:24:33.455 & -00:02:21.73 & 2.3330          & 0.1$\pm$0.05    & 2.554 (H$\alpha$) / 5.259 ([O{\sc ii}]) & $>$78 \\
2          & 02:24:33.591 & -00:02:24.22 & 2.2455          & 0.06$\pm$0.03   & 2.421 (H$\alpha$) / 5.024 ([O{\sc ii}]) & $>$47  \\
3          & 02:24:33.708 & -00:02:25.71 & 1.6740          & 0.04$\pm$0.02   & 1.550 (H$\alpha$) / 3.159 ([O{\sc ii}]) & $>$31  \\
4          & 02:24:33.315 & -00:02:23.28 & 2.4302          & 0.03$\pm$0.02   & 2.702 (H$\alpha$) / 5.520 ([O{\sc ii}]) & $>$23 \\
\hline\hline
\label{table:serendip_sources}
\end{tabular}
\caption{Properties of the serendipitous sources in the VIMOS and SINFONI IFU fields of view}
}
\end{center}
\end{table*}

\noindent
\smallskip
{\bf Near-Infrared}\\
\noindent
{\it SINFONI Source 1:} The proximity of this source to the z=4.88
critical line makes it more likely to be [O{\sc ii}] for z=5.25.  We
can not rule out H$\alpha$ at $z$=2.55, although the lack of [O{\sc
  iii}] or H$\beta$ at 1.8$\mu$m (although near to the atmospheric
$H$-band absorption) or [N{\sc ii}]$\lambda$6583 emission makes it more
likely [O{\sc ii}].  \smallskip

\noindent {\it SINFONI Source 2:} This source lies close to the high
amplification lines for $z$=2-3 and therefore we suggest H$\alpha$ is
the most likely line identification.  There is no strong [O{\sc iii}]
in the $H$-band.  \smallskip

\noindent
{\it SINFONI Source 3:} The proximity to the z$\sim$1.5 critical lines
makes it more likely to be H$\alpha$ than high redshift [O{\sc ii}]
\smallskip

\noindent {\it SINFONI Source 4:} This is the weakest emission line in the
sample, although it still fulfils the selection criterion.  The most
likely line identification is [O{\sc ii}] at $z$=5.51 due to its
proximity to the $z\sim$5 critical curve.  This emission line could
also be identified at H$\alpha$ for z=2.70, however, we note that there
are no signs of strong [O{\sc iii}] with a flux ratio limit of [O{\sc
  iii}]/H$\alpha$ emission line ratio $\lsim0.2$ \smallskip


\section{Spectroscopic Properties of the Cluster Galaxies}

Table~\ref{table:cluster_members} gives the spectroscopic properties
(RA, Dec, [O{\sc ii}] emission line flux, line width (FWHM) and
Equivalent Width (EW)) of the cluster galaxies.

\begin{table*}
\begin{center}
{\centerline{\sc Table B1.}}
{\centerline{\sc Spectroscopic Properties of the Cluster Galaxies}}
\begin{tabular}{lccccccc}
\hline
\hline
\noalign{\smallskip}
ID  & RA           & Dec          & $z$    & Features          & [O{\sc ii}] Flux                    & [O{\sc ii}] FWHM &  EW    \\ 
    &              &              &        &                   & ($\times$10$^{-17}$erg/cm$^{2}$/s)) & (\AA)            & (\AA)  \\
\hline
\hspace{-0.3cm}\\
CG1 & 02:24:34.097 & -00:02:30.90 & 0.7784 & H\&K              &  -           & -    &      \\
CG2 & 02:24:34.246 & -00:02:32.54 & 0.7725 & H\&K              &  -           & -    &      \\
CG3 & 02:24:34.184 & -00:02:31.29 & 0.7603 & [O{\sc ii}]       & 8.5$\pm$1.0  &  5.7 & 7.0  \\
CG4 & 02:24:34.717 & -00:02:35.22 & 0.7845 & [O{\sc ii}]       & 12.0$\pm$1.0 &  7.0 & 8.2  \\
CG5 & 02:24:34.751 & -00:02:28.75 & 0.7864 & H\&K, [O{\sc ii}] & 10.0$\pm$2.0 &  4.1 & 22.0 \\
CG6 & 02:24:34.472 & -00:02:27.71 & 0.7848 & H\&K, [O{\sc ii}] & 20.0$\pm$1.0 & 11.1 & 17.0 \\
\hline
\hline
\label{table:cluster_members}
\end{tabular}
\caption{Note: Properties of the brightest cluster galaxies, as
  measured from our VIMOS IFU observations.  Equivalent widths (EW) and
  emission line FWHM are given in the rest-frame of the galaxy.}
\end{center}
\end{table*}

\end{document}